\documentclass[preprint,3p,11pt,sort&compress]{elsarticle}

\usepackage{amsmath,amsthm,amssymb}
\usepackage{MnSymbol}
\usepackage{graphicx}
\usepackage[usenames,dvipsnames]{color}
\usepackage{url}
\usepackage[mathlines]{lineno}

\newtheorem{theorem}{Theorem}
\newtheorem{proposition}[theorem]{Proposition}
\newtheorem{lemma}[theorem]{Lemma}
\newtheorem{conjecture}[theorem]{Conjecture}
\newtheorem{corollary}[theorem]{Corollary}
\newtheorem{definition}[theorem]{Definition}
\newtheorem{example}[theorem]{Example}

\newtheorem{remark}[theorem]{Remark}

\newcommand{\C}{\ensuremath{\textbf{C}^{\infty}}}

\newcommand{\Kola}{\mathcal{K}}
\newcommand{\W}{\mathcal{W}}
\newcommand{\GammaS}{\Gamma_{\! s}}
\newcommand{\GammaD}{\Gamma_{\! d}}

\newcommand{\h}{\textit{height}}

\newcommand{\figref}[1]{\figurename~\ref{#1}}
\renewcommand{\epsilon}{\varepsilon}

\begin{document}

\sloppy 

\begin{frontmatter}

\title{Vertical Representation of \C-words\tnoteref{t1}}
\tnotetext[t1]{Some of the results in this paper were presented at the \emph{Conference for the 20th Anniversary of LaCIM}, 29-31 August 2010, Montreal, Canada \cite{FeFi10b}. The second author has been partially funded by ANR EMC (ANR-09-BLAN-0164).}

\journal{Theoretical Computer Science}

\author[nice]{Jean-Marc F\'{e}dou}
 \ead{fedou@i3s.unice.fr}

\author[palermo]{Gabriele Fici\corref{cor1}}
\ead{gabriele.fici@unipa.it}

\address[nice]{Laboratoire d'Informatique, Signaux et Syst\`emes de Sophia-Antipolis \\ CNRS \& Universit\'e Nice Sophia Antipolis\\ 2000, Route des Lucioles - 06903 Sophia Antipolis cedex, France}

\address[palermo]{Dipartimento di Matematica e Informatica \\ Universit\`a di Palermo\\ Via Archirafi 34 - 90123 Palermo, Italy}

\cortext[cor1]{Corresponding author.}

\begin{abstract}
We present a new framework for dealing with \C-words, based on their left and right frontiers. This allows us to give a compact representation of them, and to describe the set of \C-words through an infinite directed acyclic graph $G$. This graph is defined by a map acting on the frontiers of \C-words. We show that this map can be defined recursively and with no explicit reference to \C-words. We then show that some important conjectures on \C-words follow from analogous statements on the structure of the graph $G$. 
\end{abstract}

\begin{keyword}
 Kolakoski word, $\C$-words, directed acyclic graph, recursive function, directed set.
\end{keyword}
\end{frontmatter}

\section{Introduction}\label{sec:intro}

Every finite or infinite word $w$ over a finite alphabet $\Sigma$ of cardinality greater than $1$ can be written in a unique way by replacing maximal blocks of consecutive identical letters, called runs, with the single letter having the length of the block as exponent, as in $a^{2}b^{1}c^{3}=aabccc$. The sequence of exponents of $w$ is called the \emph{exponent word} of $w$, and is denoted by $\Delta(w)$. 

Fixed an integer alphabet $\Sigma$ (i.e., a finite subset of $\mathbb{N}\setminus \{0\}$), the words over $\Sigma$ such that their exponent word is still a word over $\Sigma$ are called \emph{differentiable words}. 

Studying the exponent words in the context of symbolic dynamics, Oldenburger \cite{Ol}  showed in 1938 that there exist infinite words  that coincide with their exponent word, and that these must be non-periodic. For example, if $\Sigma=\{1,2\}$ there exist precisely two such words, namely the word
$$\Kola=221121221221121122121121\cdots$$ and the word $1\cdot \Kola$.

The word $\Kola$ is known as the \emph{Kolakoski word} \cite{Ko65}, although perhaps it should be  more properly called the \emph{Oldenburger-Kolakoski word}.

Several longstanding conjectures on the combinatorial structure of the Kolakoski word  remain unproved---some of them are in the original 1938 paper by Oldenburger. For example, in its paper Oldenburger asks whether or not there exist recurrent\footnote{Recall that a word is said to be recurrent if all of its factors appear infinitely often.} words  that coincide with their exponent word. This question has been answered in the affirmative for  words over binary alphabets in which the two letters have the same parity \cite{BJP08}, but it is still open for words over the alphabet $\{1,2\}$ (see \cite{Ki79}), and in particular thus for the Kolakoski word. It is easy to see that a sufficient condition for the Kolakoski word being recurrent is that its set of factors is closed under complement (swapping of $1$'s and $2$'s) (see \cite{De80}).
In fact, Brlek and Ladouceur \cite{BrLa03} proved that it is even sufficient to prove that  $\Kola$ contains arbitrarily long palindromic factors. However, all these properties are still unproven.

This motivates us to study the set of factors of the Kolakoski word.

\subsection{The set of $\C$-words}

In order to study the finite factors of the Kolakoski word, the operator $\Delta$ is not convenient, since it does not preserve the set of factors. For example, $121$ is a factor of $\Kola$ but $\Delta(121)=111$ is not. For this reason, we use the operator $D$, called \emph{the derivative} in \cite{De80} but, as an anonymous referee pointed out, yet introduced in \cite{Ol} under the name of \emph{proper exponent block},
 that consists in discarding the first and/or the last letter in $\Delta(w)$ if these are equal to $1$. For example, the derivative of $121$ is $D(121)=1$, while the derivative of $12$ is $\epsilon$, the empty word. The set of finite words over $\Sigma=\{1,2\}$ that are derivable arbitrarily many times over $\Sigma=\{1,2\}$ is called \emph{the set of $\C$-words}. It is  closed under complement and reversal, and contains the set of factors of the Kolakoski word. Thus, one of the most important open problems about the Kolakoski word is to decide  whether all the $\C$-words occur as factors in $\Kola$:

\begin{conjecture}\cite{De80}\label{conj:kola}
Every $\C$-word is a factor of the Kolakoski word.
\end{conjecture}

Actually, the set of $\C$-words contains the set of factors of any right-infinite word over $\{1,2\}$ having the property that an arbitrary number of applications of $\Delta$ still produces a word over $\{1,2\}$. Such words are called \emph{smooth words} \cite{BrLa03,BeBrCho05}. 
Nevertheless, the existence of a smooth word such that the set of its factors is equal to the whole set $\C$ is still an open question. 
By the way, we notice that should Conjecture \ref{conj:kola} be true, the Kolakoski word would be recurrent (see \cite{De80}).

In addition to the aforementioned problems, there is a conjecture of Keane \cite{Ke91} stating that the frequencies of $1$s and $2$s in the Kolakoski word exist and are equal to $1/2$. Chv\'atal \cite{Ch93} showed that if these limits exist, they are very close to $1/2$ (actually, between $0.499162$ and $0.500838$).

Up to now, only few combinatorial properties of $\C$-words have been established. Weakley \cite{We89} started a classification of $\C$-words and obtained significant results on their complexity function. Carpi \cite{Ca94} proved that the set $\C$ contains only a finite number of squares, and does not contain cubes (see also \cite{Le94} and \cite{BrduLaVu06}). This result generalizes to repetitions with gap, i.e., to the \C-words of the form $uzu$, for a non-empty $z$. Indeed, Carpi \cite{Ca94} proved that for every $k>0$, only finitely many $\C$-words of the form $uzu$ exist with $z$ not longer than $k$. In a previous paper \cite{FeFi12}, we proved that for any \C-word $u$, there exists a $z$ such that $uzu$ is a \C-word, and $|uzu|\leq C|u|^{2.72}$, for a suitable constant $C$.
In the same paper, we proposed the following conjecture:

\begin{conjecture}\cite{FeFi12}\label{conj:univ}
For any $u,v$ \C-words, there exists $z$ such that $uzv$ is a \C-word.
\end{conjecture}

Despite Conjecture \ref{conj:univ} being a weaker condition than Conjecture \ref{conj:kola}, it remains an open question. 

\subsection{Outline of the results}

We find convenient to represent \C-words together with all their non-empty derivatives, as shown in \figref{fig:der} (we adopt the convention that $w=D^{0}(w)$ for any word $w$). 

\begin{figure}[ht]
 \[\begin{tabular}{p{15mm}p{15mm}p{27mm}p{20mm}}
\hline \rule[-6pt]{0pt}{18pt} $D^{0}$ & $2121122$   & $21221211221221$ & $2122121122$    \\
 \rule[-6pt]{0pt}{13pt} $D^{1}$ & $1122$       &  $12112212$ & $121122$     \\
 \rule[-6pt]{0pt}{15pt} $D^{2}$ & $22$     & $1221$  &$122$       \\
 \rule[-6pt]{0pt}{15pt} $D^{3}$ & $2$     & $2$  &$2$       \\
 \hline  \rule[-1pt]{0pt}{1pt}
\end{tabular} \] \caption{The $\C$-words $u=2121122$, $u'=21221211221221$ and $u''=2122121122$, represented together with their non-empty derivatives.\label{fig:der}}
\end{figure} 

Notice that every $\C$-word $w$ can be reduced to the empty word with a finite number $k$ of applications of the derivative, and we call the least of such $k$ the \emph{height} of $w$. For example, the words in \figref{fig:der}  all have height $3$.

The sequence of the first letters of the non-empty derivatives of a $\C$-word $w$ can be encoded into a word $\Psi(w)$ over the alphabet $\Sigma_{0}=\{0,1,2\}$, that we call the \emph{left frontier} of $w$. For every $0\leq i<k$, the $(i+1)$th letter of $\Psi(w)$ is $0$ if $D^{i-1}(w)$ begins in $122$ or $211$, or the first letter of the $i$th derivative $D^{i}(w)$ otherwise. 

Analogously, one can define the \emph{right frontier} of $w$, $\Psi^{R}(w)$, as the left frontier of the reversal of $w$. The pair $[\Psi(w),\Psi^{R}(w)]$ is called the \emph{vertical representation} of the word $w$, and allows one to uniquely represent  any $\C$-word by means of a pair of words whose length is logarithmic in the length of the \C-word (Theorem \ref{teor:vertical}).

For example, the vertical representations of the words $u$, $u'$ and $u''$ in \figref{fig:der} are, respectively, $[2122,2222]$, $[2110,1010]$ and $[2110,2222]$.

The map $\Psi$ induces an equivalence relation on the set of \C-words defined by the property of having the same left frontier. We call this equivalence the \emph{$\Psi$-equivalence} and its classes \emph{$\Psi-classes$}. We show that two words have the same left frontier if and only if they have the same height and one is a prefix of the other (Theorem \ref{theor:Psi}). Therefore, the words belonging to the same $\Psi$-class form a prefix chain.

The words that do not have any $0$ in their vertical representation are called \emph{minimal}, because each derivative is obtained from a primitive of minimal length. Thus, for any word $U\in \Sigma^{*}$, there exists a unique single-rooted (i.e., with last non-empty derivative having length one) minimal word having left frontier $U$, and we set $\GammaS(U)$ the right frontier of this word. Analogously, there exists a unique double-rooted (i.e., with last non-empty derivative having length $2$) minimal word having left frontier $U$, and we set $\GammaD(U)$ its right frontier. The maps $\GammaS$ and $\GammaD$ can be naturally extended to any word $U$ over the alphabet $\Sigma_{0}$, by taking $\GammaS(U)$ (resp.\ $\GammaD(U)$) as the right frontier of the shortest single-rooted (resp.\ double-rooted) \C-word having left frontier $U$.

For example, given the word $U=2122$, the single-rooted minimal word having left frontier $U$ is the word $u=2121122$ (see \figref{fig:der}), and therefore $\GammaS(2122)=2222$. 
The word $u''=2122121122$ in \figref{fig:der} is the shortest $\C$-word having left frontier $U'=2110$, so we have $\GammaS(U')=\GammaS(U)=2222$. The reader can check that the minimal double-rooted word having left frontier $U$ is the \C-word $v=212112212211$ whose right frontier is $\Psi^{R}(v)=\GammaD(U)=1221$, and that the shortest double rooted  \C-word having left frontier $2110$ is $v''=212212112212211=212\cdot v$, so that $\GammaD(2110)=\Psi^{R}(v'')=1221$.

Thus, we can define an equivalence relation, called \emph{minimal equivalence}, on the set of $\C$-words, by setting $u\equiv_{\GammaS}v$ if and only if $\GammaS\Psi(u)=\GammaS\Psi(v)$, where $\GammaS$ and $\Psi$ are composed in the usual way. The equivalence classes of this relation, called \emph{minimal classes}, allow one to reduce the study of the properties of \C-words to these of single-rooted minimal words, since there is exactly one single-rooted minimal word in each minimal class---and it is in fact the shortest word in the class. Furthermore,  any minimal class is the union of $\Psi$-classes, and we show that two $\Psi$-classes belong to the same minimal class if and only if their shortest words have the same height and one is a suffix of the other (Theorem \ref{theor:Gamma}). 

The minimal classes can be represented over an infinite directed acyclic graph $G$, whose set of nodes is $\Sigma^{*}$ and for every non-empty node $U$ there are three outgoing edges: one with label $1$ that goes to node $U1$; one with label $2$ that goes to node $U2$; and finally one with label $0$ that goes to node $\Theta(U)2$, where $\Theta$ is the composition map $\GammaS\GammaD$---the empty node has only the outgoing edges labeled by $1$ and $2$ respectively. 
In fact, we prove that the minimal class of the word $U0$ is the word $\Theta(U)2$ (Theorem \ref{theor:Theta}). Hence, a minimal class $U$ can be extended into three minimal classes of greater height, namely the classes $U2$, $U1$ and $\Theta(U)2$. Thus, in the graph $G$, for any $U\in \Sigma^{*}$, the $\Psi$-classes forming the minimal class $U$ are precisely the labels of the paths from the root $\epsilon$ to the node $U$.

The graph $G$ is therefore an infinite complete binary tree with additional edges defined by the map $\Theta$, i.e., $G$ is completely defined by the action of $\Theta$. We prove that the maps $\GammaS$ and $\GammaD$, and therefore the map $\Theta$, can be defined recursively and independently from the context of $\C$-words (Theorem \ref{theor:recursive}). Thus, also the graph $G$ can be defined recursively and independently from the context of $\C$-words. This result may open new perspectives in the study of \C-words. As an example, we formulate two conjectures on the graph $G$ that, if proven, would imply the validity of some important conjectures on the structures of $\C$-words and on the Kolakoski word.

\bigskip

The paper is organized as follows. In Section \ref{sec:nota} we introduce \C-words and their combinatorial properties; in Section \ref{sec:vertical} we introduce the vertical representation of \C-words; then, in Section \ref{sec:frontiers} we define the maps $\GammaS$ and $\GammaD$, and the minimal classes, and we present the graph $G$ of minimal classes. 
In Section \ref{sec:graph}  we deal with the extensions on the right of \C-words from the point of view of their left frontiers, and introduce the graph $G$.
In Section \ref{sec:recursive} we give recursive formulae for the maps acting on the frontiers, leading to a recursive definition of the graph $G$. Finally, in Section \ref{sec:final}, we discuss conclusions and final remarks.

\section{Preliminaries}\label{sec:nota}

We fix the two-letter alphabet $\Sigma=\{1,2\}$, and we call its elements \emph{letters}.  A \emph{word} over $\Sigma$ is a finite sequence of letters from $\Sigma$. The \emph{length} of a word $w$ is denoted by $|w|$. The \emph{empty word} has length zero and is denoted by $\epsilon$. The set of all words over $\Sigma$ is denoted by $\Sigma^*$. The set of all words over $\Sigma$ having length $n$ is denoted by $\Sigma^n$. 

Let $w\in \Sigma^{*}$. If $w=uzv$ for some $u,z,v\in\Sigma^{*}$, we say that $z$ is a \emph{factor} of $w$. In the case where $u=\epsilon$ (resp.\ $v=\epsilon$), $z$ is called a \emph{prefix} of $w$ (resp.\ a \emph{suffix} of $w$). By definition, a word is a factor (resp.\ a prefix, resp.\ a suffix) of itself. Therefore, we call $z$ a \emph{proper} factor (resp.\ prefix, resp.\ suffix) of $w$ if $z$ is different from $w$.

The \emph{reversal} (or \emph{mirror image}) of a word $w$ is the word $\widetilde{w}$ obtained by writing the letters of $w$ in the reverse order. For example, the reversal of $w=11212$ is $\widetilde{w}=21211$.
The \emph{complement} of a word $w$ is the word $\overline{w}$ obtained by swapping the letters of $w$, i.e., by changing the $1$s in $2$s and the $2$s in $1$s. For example, the complement of $w=11212$ is $\overline{w}=22121$. We also set $\widetilde{\epsilon}=\overline{\epsilon}=\epsilon$.

A \emph{right-infinite word} over $\Sigma$ is an unending sequence of letters from $\Sigma$. The set of all right-infinite words over $\Sigma$ is denoted by $\Sigma^{\omega}$.

Let $w$ be a finite word over an alphabet $\Sigma$. Then $w$ can be uniquely written as a concatenation of maximal blocks of identical letters (called \emph{runs}), i.e., $w=x_1^{i_1}x_2^{i_2}\cdots x_n^{i_n}$, with $x_{j}\in \Sigma$, $x_{j}\neq x_{j+1}$, and $i_{j}>0$. The \emph{exponent word} (also known as \emph{run-length encoding}) of $w$, denoted $\Delta(w)$, is the sequence of exponents $i_{j}$, i.e., one has $\Delta(w)=i_1i_2\cdots i_n$. These definitions extend naturally to right-infinite words.

\begin{definition}\cite{BrLa03}
A right-infinite word $\W\in \Sigma^{\omega}$ is called a \emph{smooth word over $\Sigma$} if for every integer $k> 0$ one has $\Delta^k(\W)\in \Sigma^{\omega}$. 
\end{definition}

In this paper we focus on the set of factors of smooth words, called \C-words. We start by recalling some definitions.

\begin{definition}\cite{De80}
A word $w\in \Sigma^{*}$ is \emph{differentiable} if $\Delta(w)$ is still a word over $\Sigma$. 
\end{definition}

  Since $\Sigma=\{1,2\}$, we have that $w$ is differentiable if and only if neither $111$ nor $222$ appear in $w$. 

\begin{definition}\cite{De80}
The \emph{derivative} is the map $D$ defined on differentiable words by:
$$D(w) = \left\{ \begin{array}{lllll}
\epsilon & \mbox{if $\Delta(w)=1$ or $w=\epsilon$,}\\
\Delta(w) & \mbox{if $\Delta(w)=2x2$ or $\Delta(w)=2$,}\\
x2 & \mbox{if $\Delta(w)=1x2$,}\\
2x & \mbox{if $\Delta(w)=2x1$,}\\
x & \mbox{if $\Delta(w)=1x1$.}
\end{array} \right.$$
\end{definition}

In other words, the derivative $D(w)$ is obtained from $\Delta(w)$ by erasing the first and/or the last letter if they are equal to one. 

Let $k\geq 0$. A word $w$ is \emph{$k$-differentiable} on $\Sigma$ if $D^k(w)$ is defined. Of course, a word $w$ is $k$-differentiable if and only if for every $0\leq j<k$ the word $D^{j}(w)$ does not contain $111$ nor $222$ as a factor. We adopt the convention that $D^0(w)=w$. 
 Clearly, if a word is $k$-differentiable, then it is also $j$-differentiable for every $0\leq j \leq k$. 
 
 A \C-word is a word that is differentiable arbitrarily many times, that is to say, a word that can be reduced to the empty word with finitely many derivations. As a direct consequence of the definitions, the set of \C-words is  closed under reversal and complement.
 
\begin{definition}\cite{We89} 
The \emph{height} of a \C-word $w$ is the least integer $k$ such that $D^{k}(w)=\epsilon$. It is denoted by $\h(w)$.
\end{definition}

\begin{definition}\cite{FeFi12} 
Let $w$ be a \C-word of height $k>0$. The \emph{root} of $w$ is $D^{k-1}(w)$. Therefore, the root of $w$ belongs to $\{1,2,12,21\}$. Consequently, $w$ is called \emph{single-rooted} if its root has length one or \emph{double-rooted} if its root has length two.
\end{definition}

\begin{definition}\cite{De80} 
A \emph{primitive} of a word $w$ is any word $w'$ such that $D(w')=w$.  
\end{definition}

It is easy to see that any $\C$-word has two, four or eight different primitives (actually, it has two primitives if it starts and ends with 1, eight primitives if it starts and ends with 2, and four primitives otherwise). For example, the word $22$ has eight primitives ($1122$, $21122$, $11221$, $211221$, $2211$, $12211$, $22112$, $122112$), while the word $121$ has only two primitives ($121121$, $212212$). The two primitives of minimal (resp.\ maximal) length are called \emph{short primitives} (resp.\ \emph{long primitives}) \cite{We89}.

\begin{definition}\cite{FeFi12} 
Let $w$ be a  $\C$-word of height $k>1$. We say that $w$ is a \emph{minimal word} (resp.\ \emph{a maximal word}) if for every $0\leq j\leq k-2$, $D^{j}(w)$ is a short (resp.\ long) primitive of $D^{j+1}(w)$. The words of height $1$ are assumed to be both minimal and maximal.
\end{definition}

So, minimal words are those \C-words in which every derivative is obtained from a short primitive, while maximal words are those in which every derivative is obtained from a long primitive.

\begin{definition}\cite{FeFi12} 
A \C-word $w$ is \emph{left minimal} (resp.\ \emph{left maximal}) if it is a prefix of a minimal (resp.\ maximal) word.
Analogously, $w$ is \emph{right minimal} (resp.\ \emph{right maximal}) if it is a suffix of a minimal (resp.\ maximal) word.
\end{definition}

Clearly, a word is minimal (resp.\ maximal) if and only if it is both left minimal and right minimal (resp.\ left maximal and right maximal).

\begin{example}[see \figref{fig:der}]\label{ex:minimal}
 The word $u=2121122$ is minimal, since all the derivatives come from primitives of minimal length; the word $u'=21221211221221=212\cdot u \cdot 1221$ is maximal, since all the derivatives come from primitives of maximal length; the word $u''=2122121122=212\cdot u$ is left maximal and right minimal. Notice that 
 the three words have same height and same root, and that $u$ is a suffix of $u''$, which in turn is a prefix of $u'$.
\end{example}

Weakley \cite{We89} started a classification of $\C$-words based on extendibility. Indeed, any $\C$-word has arbitrarily long left and right extensions. That is to say, for any $\C$-word $w$ at least one between $1w$ and $2w$  is a $\C$-word, and, analogously, at least one between $w1$ and $w2$ is a $\C$-word.

\begin{definition}\cite{We89}
A \C-word $w$ is \emph{left doubly extendible} (LDE) if $1w$ and $2w$ are both $\C$-words; otherwise, $w$ is \emph{left simply extendible} (LSE). Analogously, $w$ is \emph{right doubly extendible} (RDE) if $w1$ and $w2$  are both $\C$-words; otherwise, $w$ is \emph{right simply extendible} (RSE). A \C-word $w$ is \emph{fully extendible} (FE) if $1w1$, $1w2$, $2w1$ and $2w2$ are all \C-words. 
\end{definition}

Based on the previous definitions and on a result of Weakley (\cite{We89}, Proposition 3) one can establish the following structural result (see also \cite{FeFi12}).

\begin{theorem}\label{theor:Weakley}
Let $w$ be a \C-word. The following conditions are equivalent:
\begin{enumerate}
\item $w$ is FE (resp.\ $w$ is LDE, resp.\ $w$ is RDE).
\item $w$ is double-rooted maximal (resp.\ $w$ is left maximal, resp.\ $w$ is right maximal).
\item $w$ and all its non-empty derivatives (resp.\ $w$  and all its derivatives longer than one) begin with two different letters and end with two different letters (resp.\ begin with two different letters, resp.\ end with two different letters).
\end{enumerate}
\end{theorem}

It is worth noticing that a \C-word $w$ can be both LDE and RDE but not FE. By Theorem \ref{theor:Weakley}, this happens if and only if $w$ is single-rooted maximal.

\section{Vertical representation of \C-words}\label{sec:vertical}

In what follows, the $(i+1)$th letter of a word $w$ is denoted by $w[i]$.  So, we write a word $w$ of length $n>0$ as $w = w[0]w[1]\cdots w[n-1]$.

Given a \C-word $w$, the sequence obtained by concatenating the first letter of each derivative, and that obtained by concatenating the last letter of each derivative, form a pair of words that represent $w$. Unfortunately, this representation is not injective. Take for example $w=2211$ and $w'=21121221$, for both words this pair is equal to $(222,122)$. 

In order to obtain an injective representation of \C-words, we consider the alphabet  $\Sigma_{0}=\{0,1,2\}$ and recall the following definition, given in \cite{FeFi12}.

\begin{definition}\label{def:frontier}
Let $w$ be a $\C$-word of height $k>0$. The \emph{left frontier} of $w$ is the word $\Psi(w)$  of length $k$  over the alphabet $\Sigma_{0}$ defined by: $\Psi(w)[0]=w[0]$ and, for $0<i<k$,
 $$\Psi(w)[i] = \left\{ \begin{array}{lllll}
0 & \mbox{if $D^{i-1}(w)$ begins in $122$ or $211$,}\\
D^{i}(w)[0]  & \mbox{otherwise.}
\end{array} \right.$$
For the empty word, we set $\Psi(\varepsilon)=\varepsilon$.
The \emph{right frontier} of $w$ is the word $\Psi^{R}(w)=\Psi(\widetilde{w})$, i.e., the left frontier of the reversal of $w$. The \emph{vertical representation} of $w$ is the pair $[\Psi(w),\Psi^{R}(w)]$. 
\end{definition}

In other words, to obtain the left (resp.\ the right) frontier of $w$, one has to take the first (resp.\ the last) letter of each derivative of $w$ and replace a $2$ with a $0$ whenever the primitive above is not left minimal (resp.\ not right minimal).

\begin{example}\label{ex:vert}
Consider the $\C$-word $w=21221211221$, whose derivatives are listed below. 
\[ \begin{tabular}{p{15mm} l}
\hline \rule[-6pt]{0pt}{18pt}$D^{0}$  & $21221211221$ \\
\rule[-6pt]{0pt}{13pt}$D^{1}$  & $121122$ \\
\rule[-6pt]{0pt}{13pt}$D^{2}$  & $122$\\
\rule[-6pt]{0pt}{13pt}$D^{3}$  & $2$\\
\hline \rule[-2pt]{0pt}{2pt}
\end{tabular}  \]
The word $D^{2}(w)=122$ is not a left minimal primitive of the word $D^{3}(w)=2$, and therefore $\Psi(w)[3]$, the fourth letter of the left frontier of $w$, equals $0$; analogously, the word $w=21221211221$ is not a right minimal primitive of $D(w)=121122$, and therefore $\Psi^{R}(w)[1]$, the second letter of the right frontier of $w$, equals $0$. Hence, the vertical representation of $w$ is $[2110,1022]$.
\end{example}

In \cite{FeFi12}, we established the following result (of which we include a proof for the sake of completeness).

\begin{theorem}\label{teor:vertical}
Any $\C$-word  is uniquely determined by its vertical representation, i.e., the map $[\Psi,\Psi^{R}]:\C \mapsto \Sigma_{0}^{*} \times \Sigma_{0}^{*}$  is injective. 
\end{theorem}

\begin{proof}
 By induction on the height. A rapid check shows that all the \C-words of height $1$ have  different vertical representations. Suppose the statement true for any integer smaller than $k>1$ and let $w,w'$ be two different \C-words of height $k$. Suppose by contradiction that $w$ and $w'$ have the same vertical representation. Then $D(w)$ and $D(w')$ also have the same vertical representation. By induction hypothesis we must have $D(w)=D(w')$. Therefore, $w$ and $w'$ are two different primitives of the same \C-word, and, by Definition \ref{def:frontier},  $w$ and $w'$ cannot have the same vertical representation.
\end{proof}

 In what follows, uppercase letters ($U,V,W,\ldots$) will denote vertical words, i.e., words over $\Sigma_{0}$ whose first letter is different from $0$, coding the (left or right) frontier of a $\C$-word; lowercase letters ($u,v,w,\ldots$) will still denote $\C$-words.

It is worth noticing that given any two vertical words $U,V$ of the same length, it may happen that no \C-word exists having vertical representation $[U,V]$. For example, no $\C$-word exists having vertical representation $[11,21]$. 
An interesting problem would be that of determining, given two words $U$ and $V$, whether there exists a \C-word having vertical representation $[U,V]$. We leave this as an open problem.

\begin{lemma}\label{lem:ext}
 Let $u$ be a right maximal (equivalently, an RDE) word and let $U=\Psi(u)$ be its left frontier. Then $u1$ and $u2$ are right minimal words, and:
 
\begin{itemize}
 \item if $u$ is single-rooted, then $\{\Psi(u1),\Psi(u2)\}=\{U,U2\}$;
 \item if $u$ is double-rooted, then  $\{\Psi(u1),\Psi(u2)\}=\{U1,U0\}$. 
\end{itemize}
\end{lemma}

\begin{proof}
The fact that $u1$ and $u2$ are right minimal words is a consequence of Theorem \ref{theor:Weakley}. Indeed, if $u$ is a right maximal word, then for every $0\le i < \h(u)$, one has $|D^{i}(ux)|=|D^{i}(u)|+1$ for any letter $x\in \Sigma$. Then, since by Theorem \ref{theor:Weakley} the last two letters of any derivative of $u$ longer than $1$ are different, one has that the last three letters of any derivative of $ux$ longer than $2$ are never of the form $\alpha\alpha\overline{\alpha}$, $\alpha\in \Sigma$, and this is equivalent to say that $ux$ is right minimal.

If the root of $u$ is $\alpha\in \Sigma$, then the $(\h(u)-1)$th derivative of the words $u1$ and $u2$ is $\alpha\overline{\alpha}$ or $\alpha\alpha$, and hence the first claim follows.
If instead the root of $u$ is $\alpha\overline{\alpha}$, $\alpha\in \Sigma$, then the $(\h(u)-1)$th derivative of the words $u1$ and $u2$ is $\alpha\overline{\alpha}\alpha$ or $\alpha\overline{\alpha}\overline{\alpha}$, and the second claim follows.
\end{proof}

\begin{example}
 The \C-word $w=21211221221$ is a single-rooted right maximal word. Its left frontier is $U=2122$. The word $w1$ has left frontier $U$ (it is the double-rooted minimal word in \figref{fig:Gamma}). The word $w2$, instead, has left frontier $U2$ and is therefore a single-rooted minimal word.
 
The \C-word $z=22121121$ is a double-rooted right maximal word, and its left frontier is $Z=220$. The word $z1$ has left frontier $Z0$, while the word $z2$ has left frontier $Z1$.
\end{example}

The map $\Psi$ induces an equivalence relation on the set of \C-words, defined by the property of having the same left frontier. We call this equivalence  the \emph{$\Psi$-equivalence}, and an equivalence class a \emph{$\Psi$-class}. We will denote a $\Psi$-class by the word over $\Sigma_{0}$ coding the left frontier of any \C-word in the class. 
When it is not clear from the context, we will denote by $w_{\Psi}$ the $\Psi$-class of the \C-word $w$.

\begin{example}\label{ex:psiclass}
Let $u=2121122$ be the word in \figref{fig:der}. The left frontier of $u$ is $\Psi(u)=2122$. The $\Psi$-class of $u$ is $2122_{\Psi}=\{u,u1,u12,u122,u1221,u12211,u122112,u1221121\}$.  
\end{example}

The reader can notice that in the example above the words belonging to the $\Psi$-class form a prefix chain. This is a property that holds true for any $\Psi$-class. Indeed, we have the following result:

\begin{theorem}\label{theor:Psi}
Two \C-words have the same left frontier if and only if they have the same height and one is a prefix of the other.
\end{theorem}

\begin{proof}
Suppose that $u$ and $u'$ are two words with the same left frontier. If they are not compatible with respect to the prefix order, this implies that there exists a word $w$ (the longest common prefix between $u$ and $u'$) such that $wx$ and $w\overline{x}$ have the same left frontier. But this is impossible by Lemma \ref{lem:ext}.

The other direction can be proved by a simple induction argument.
\end{proof}

\section{Minimal classes of \C-words}\label{sec:frontiers}

If $w$ is a (single-rooted or double-rooted) minimal word, then, by definition, the left and the right frontier of $w$  do not contain any $0$, i.e., are words over $\Sigma$. Moreover, since by minimality two different single-rooted (resp.\ double-rooted) minimal words cannot have the same left frontier (nor the same right frontier), the map $\Psi$ is a bijection between the set of single-rooted (resp.\ double-rooted) minimal words and $\Sigma^{*}$. That is to say, for minimal words the right frontier is determined by the left frontier and \emph{vice versa}.

So, given a non-empty word $U\in \Sigma^{*}$, we define $\GammaS(U)$ (resp.~$\GammaD(U)$) as the right frontier of the single-rooted (resp.~double-rooted) minimal word having left frontier $U$.

For the empty word, we set $\GammaS(\epsilon)=\GammaD(\epsilon)= \epsilon$.

\begin{figure}[ht]
\begin{center}
\includegraphics[height=40mm]{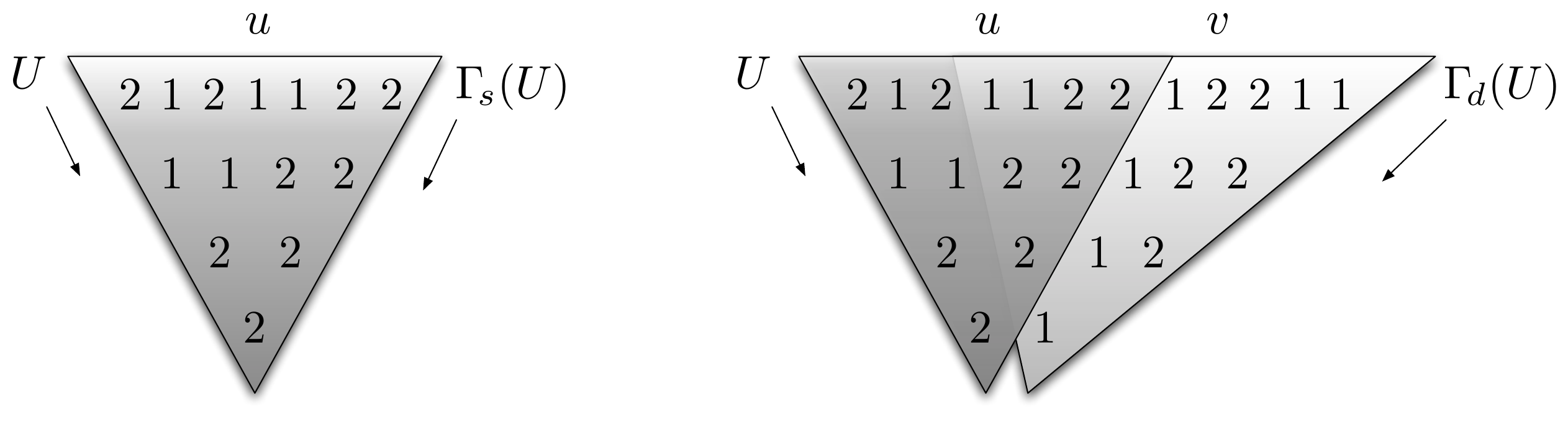}
\caption{The maps $\GammaS$ and $\GammaD$. For the vertical word $U=2122$, left frontier of the single-rooted minimal word $u=2121122$, one has $\GammaS(U)=2222$ (on the left) and $\GammaD(U)=1221$ (on the right). Notice that $\GammaD(U)$ is also the right frontier of the single-rooted minimal word $v=112212211$ (light grey, on the right).\label{fig:Gamma}}
\end{center}
\end{figure}

\begin{example}
 Let $U=2122$. The single-rooted minimal word having left frontier $U$ is $u=2121122$. The right frontier of $u$ is $2222$, therefore $\GammaS(2122)=2222$. Analogously, the double-rooted minimal word having left frontier $U$ is $u'=212112212211$. The right frontier of $u'$ is $1221$, therefore $\GammaD(2122)=1221$. Notice that the double-rooted word $u'$ is the overlap of the two single-rooted words $u$ and $v=112212211$. An illustration is given in \figref{fig:Gamma}.
\end{example}

\begin{remark}
Let $k\geq 0$. Then $\GammaS$ and $\GammaD$ are  bijections of $\Sigma^{k}$. Moreover, $\GammaS$ and  $\GammaD$ are their own inverse.
\end{remark}

The following proposition, whose proof is straightforward, relates $\GammaS$ and $\GammaD$ with $\Psi$.

\begin{proposition}
 Let $u$ be a single-rooted minimal word. Then, $\Psi(u)=\GammaS \Psi^{R}(u)$ and $\Psi^{R}(u)=\GammaS \Psi(u)$. 
 
 Analogously, let $v$ be a double-rooted minimal word. Then, $\Psi(v)=\GammaD \Psi^{R}(v)$ and $\Psi^{R}(v)=\GammaD \Psi(v)$.
\end{proposition}

We also define the composition $\Theta=\GammaS \GammaD$. Indeed, any double-rooted minimal word is the overlap between two single-rooted minimal words of the same height, one having left frontier $U$, and the other having left frontier $\Theta(U)$ (see \figurename~\ref{fig:Theta}). In other words, $\Theta(U)$ is the vertical word $V$ such that $\GammaS(V)=\GammaD(U)$.  
Analogously, one can define the composition $\GammaD \GammaS=\Theta^{-1}$, which acts symmetrically. 

\begin{example}
Let $U=2122$. Then $\Theta(U)=\GammaS \GammaD(U)=1221$, while $\Theta^{-1}(U)=\GammaD \GammaS(U)=1121$. The situation is depicted in \figref{fig:Theta}.

\begin{figure}[ht]
\begin{center}
\includegraphics[height=38mm]{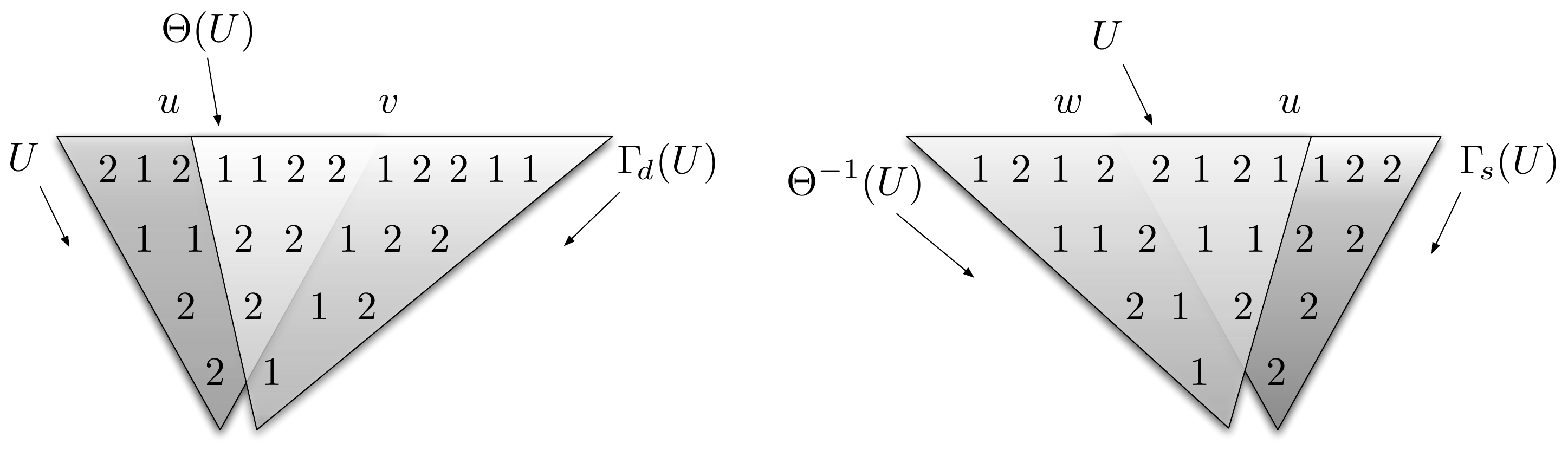}
\caption{The map $\Theta=\GammaS \GammaD$. For the vertical word $U=2122$, left frontier of the  single-rooted minimal word $u=2121122$ (dark gray,  on the left), one has $\Theta(U)=1221$, which is the left frontier of the  single-rooted minimal word $v=112212211$ (light gray,  on the left). Symmetrically, one has $\Theta^{-1}(U)=\GammaD \GammaS(U)=1121$, i.e., $\Theta^{-1}(U)$ is the left frontier of the single-rooted minimal word $w=12122121$ (light gray, on the right).\label{fig:Theta}}
\end{center}
\end{figure}
\end{example}

The maps $\GammaS$ and $\GammaD$ (and therefore $\Theta$) can be extended in a natural way to the set of words over $\Sigma_{0}$ whose first letter is different from $0$, i.e., to the $\Psi$-classes. Indeed, given a non-empty word $U$ over $\Sigma_{0}$ whose first letter is different from $0$, there exists a unique  single-rooted (resp.\ double-rooted) right minimal word having left frontier $U$, and we define $\GammaS(U)$ (resp.\ $\GammaD(U)$) as the right frontier of this word. Therefore, $\GammaS(U)$ (resp.\ $\GammaD(U)$) is the right frontier of the shortest single-rooted (resp.\ double-rooted) \C-word having left frontier $U$. Clearly, even if $U$ is a word over $\Sigma_{0}$, $\GammaS(U)$ and $\GammaD(U)$ are always words over $\Sigma$.

\begin{proposition}\label{prop:front}
 Let $U\in \Sigma_{0}^{*}$ a vertical word whose first letter is different from $0$. Then:
 
\begin{itemize}
\item the right frontier of the \C-word with vertical representation $[U2,\GammaS(U2)]$ is the complement of the right frontier of the \C-word with vertical representation $[U,\GammaD(U)]$ with $2$ appended.
\item the right frontier of the \C-word with vertical representation $[U1,\GammaS(U1)]$ is the complement of the right frontier of the \C-word with vertical representation $[U0,\GammaS(0)]$.
\end{itemize}
\end{proposition}

\begin{proof}
The statement can be proved easily by induction on the height of $U$.
\end{proof}

The map $\GammaS$ induces an equivalence relation on the set of $\Psi$-classes defined by: $U\equiv_{\GammaS} V$ if and only if $\GammaS(U)=\GammaS(V)$.  We call this equivalence the \emph{minimal equivalence}, and an equivalence class a \emph{minimal class}. When it is not clear from the context, we will note $U_{\GammaS}$ the minimal class of the $\Psi$-class $U$.

 Notice that the choice of $\GammaD$ in place of $\GammaS$ in the definition of minimal equivalence leads to the same equivalence relation, since for any $U$ and $V$ over $\Sigma_{0}$, one has $\GammaS(U)=\GammaS(V)$ if and only if $\GammaD(U)=\GammaD(V)$.

\begin{example}
Consider the vertical word $U=220$. We have $\GammaS(U)=112$ and so the minimal class of $U$ is $220_{\GammaS}=\{V \mid \GammaS(V)=112\}=\{212,220,100\}$.
\end{example}

\begin{proposition}
 Every minimal class of height $k\ge 0$ contains exactly one element of $\Sigma^{k}$.
\end{proposition}

\begin{proof}
 Let $U_{\GammaS}$ be a minimal class of height $k$. Then $\GammaS\GammaS(U)$ is an element of $U_{\GammaS}$ and belongs to $\Sigma^{k}$. The unicity is a straightforward consequence of the definition of $\GammaS$.
\end{proof}

\begin{corollary}
  There is a bijection between the set of minimal classes and the set of single-rooted minimal words.
\end{corollary}

Hence, we will identify a minimal class with the word $U\in \Sigma^{*}$ such that the single-rooted minimal word in the class has vertical representation $[U,\GammaS(U)]$. Thus, the minimal class of a $\Psi$-class $V\in \Sigma_{0}^{*}$ is the vertical word $U=\GammaS\GammaS(V)\in \Sigma^{*}$.

The following theorem provides a useful characterization of minimal classes.

\begin{theorem}\label{theor:Gamma}
Two $\Psi$-classes $U,V\in \Sigma_{0}^{*}$ of \C-words belong to the same minimal class if and only if the shortest word in $U$ and the shortest word in $V$ have the same height and one is a suffix of the other.
\end{theorem}

\begin{proof} 
 Let $U$, $V$ be two $\Psi$-classes of \C-words, and suppose that $U\equiv_{\Gamma_{s}}V$. Clearly, this implies that the words in $U$ and $V$ have the same height, say $k$. Let $Z=\GammaS(U)=\GammaS(V)$, and let $u$ and $v$ be the \C-words with vertical representation $[U,Z]$ and $[V,Z]$ respectively, i.e., the shortest words in the classes $U$ and $V$. So, $\widetilde{u}$ and $\widetilde{v}$ have the same left frontier and, by Theorem \ref{theor:Psi}, one is a  prefix of the other. Therefore, one of $u$ and $v$ is a suffix of the other.
 
Conversely, let $U$ and $V$ be two $\Psi$-classes of \C-words and suppose that $u$, the shortest word in $U$,  is a suffix  of $v$, the shortest word in $V$, and $\h(u)=\h(v)$. Therefore, $u$ and $v$ have the same right frontier. Since the vertical representations of $u$ and $v$ are $[U,\GammaS(U)]$ and $[V,\GammaS(V)]$, respectively, one has that $\GammaS(U)=\GammaS(V)$.
\end{proof}

\begin{corollary}
Two $\Psi$-classes of \C-words belong to the same minimal class  if and only if there exists an integer $n$ such that the words in one class are obtained by deleting the first $n$ letters of the words in the other class.
\end{corollary}

\begin{theorem}\label{theor:Theta}
For any $\Psi$-class $U$ different from $\epsilon$, one has $U0\equiv_{\GammaS}\Theta(U)2$.
\end{theorem}

\begin{proof} 
 By Theorem \ref{theor:Gamma}, it is sufficient to prove that the words in the $\Psi$-class $\Theta(U)2$ are suffixes of the words in the $\Psi$-class $U0$. In particular, it is sufficient to prove that the shortest word in the $\Psi$-class $\Theta(U)2$, i.e., the single-rooted right minimal one, is a suffix of the shortest word in the $\Psi$-class $U0$. This can be easily proved by induction on the height of $U$, remembering that if a \C-word $u$ is a suffix of a \C-word $v$, then $D(u)$ is a suffix of $D(v)$.
\end{proof}

The minimal equivalence can also be viewed as an equivalence relation on the set of \C-words: for any $u,v\in \C$, one has that $u$ and $v$ are in the same minimal class if and only if $\GammaS\Psi(u)=\GammaS\Psi(v)$, where $\Gamma$ and $\Psi$ are composed in the usual way. Hence, the minimal class of a \C-word $u$ is the vertical word $U=\GammaS\GammaS\Psi(u)$. That is, for every $U\in \Sigma^{*}$, one has $U=\{u\in \C \mid \GammaS\GammaS\Psi(u)=U\}$.

For example, the minimal class of the \C-word $w=21221211221221121$ is $2122$, since $\Psi(w)=2110$ and the single-rooted minimal word having right frontier $\GammaS(2110)=2222$ is the $\C$-word $u=2121122$, which has left frontier $2122=\GammaS(2222)$. Actually, $w$ is the longest word in the minimal class $2122$ (it is in fact a double-rooted maximal word) and any word in the same class is obtained from $w$ by deleting $0$ to $3$ letters from the left and $0$ to $8$ letters from the right. This means that the minimal class $2122$ is the union of $3$ $\Psi$-classes, each containing exactly $8$ $\C$-words (of which $5$ are single-rooted and $3$ are double-rooted).

In fact, the number of $\Psi$-classes forming a minimal class $U$ is the number of ``forced'' extensions to the left (that is, such that the extension to the left with the other letter would result in a word that is not $\C$) of any $\C$-word having left frontier $U$.

\section{The graph $G$ of minimal classes}\label{sec:graph}

Let $U$ be a word over $\Sigma_{0}$ whose first letter is different from $0$. By the results in the previous section, each word in a non-empty $\Psi$-class $U$ can be extended to the right to a word of greater height in three different ways, obtaining: 

\begin{enumerate}
 \item the word with vertical representation $[U2,\GammaS(U2)]$ (but only starting from single-rooted words in $U$); 
 \item the word with vertical representation $[U1,\GammaS(U1)]$;
 \item  the word with vertical representation $[U0,\GammaS(U0)]$, that is not (left) minimal, and belongs to the minimal class $\Theta(U)2$. 
\end{enumerate}

\begin{figure}[ht]
\begin{center}
\includegraphics[height=36mm]{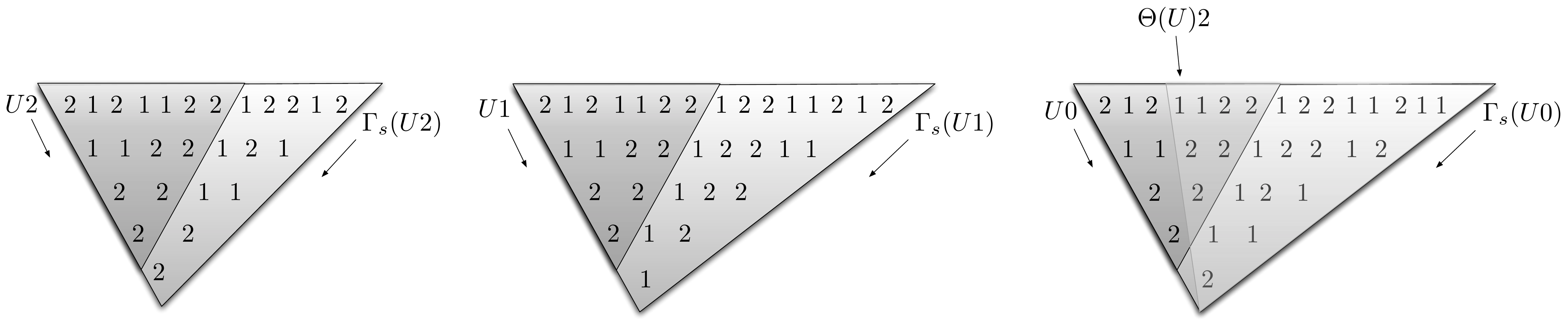}
\caption{Let $U=2122$. The words in the $\Psi$-class $U$ can be extended to the right up to a word of greater height in three ways, obtaining: the word with vertical representation $[U2,\GammaS(U2)]$ (left), the word with vertical representation $[U1,\GammaS(U1)]$ (center), and the word with vertical representation $[U0,\GammaS(U0)]$ (right), that is not minimal, and belongs to the minimal class $\Theta(U)2$.\label{fig:Extensions}}
\end{center}
\end{figure}

An example is given in \figref{fig:Extensions}. 

\begin{figure}[ht]
\begin{center}
\includegraphics[height=50mm]{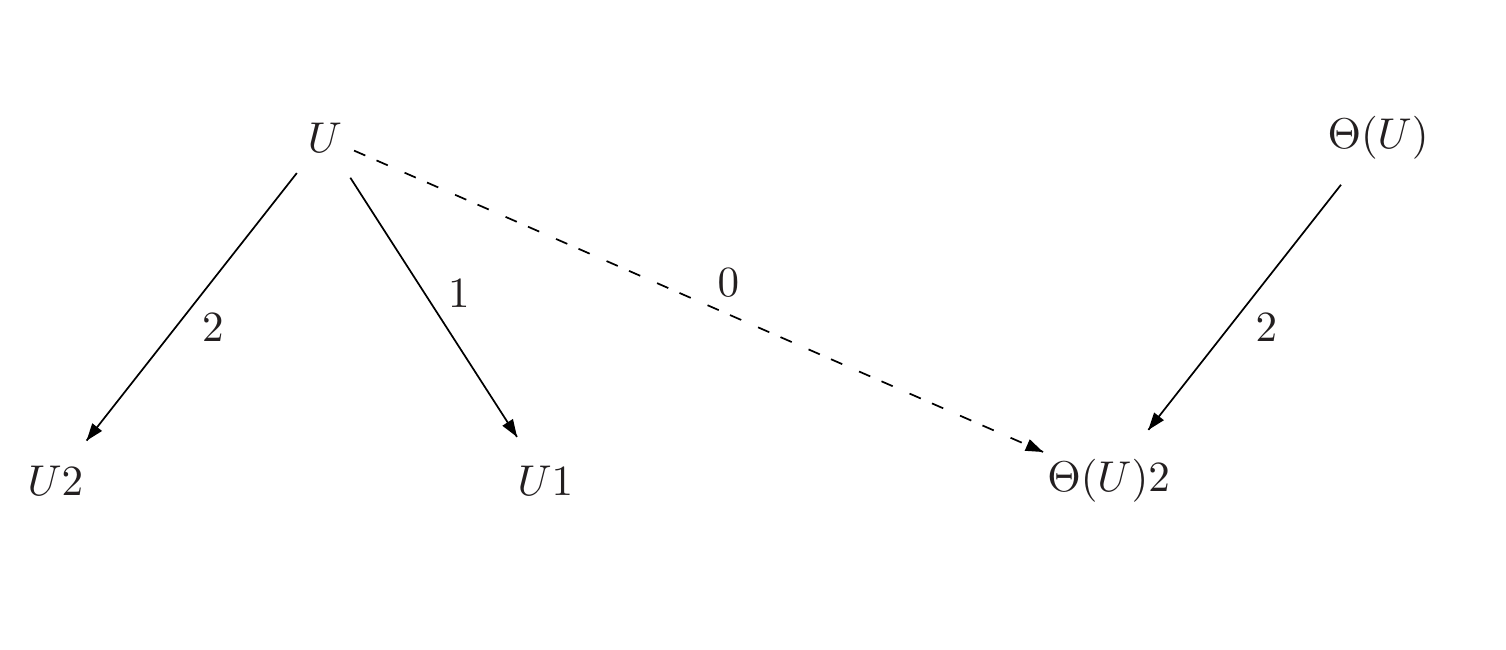}
\caption{The three extensions of a minimal class $U$ into minimal classes of greater height: $U1$, $U2$ and $\Theta(U)2$.\label{fig:Graph5}}
\end{center}
\end{figure}

In particular, if $U\in \Sigma^{*}$, we have shown how minimal classes are extended to the right into minimal classes of greater height, as illustrated in \figref{fig:Graph5}.

We can thus define an infinite directed acyclic graph $G$ for representing the minimal classes. The graph $G$ was introduced in \cite{FeFi12}. It is defined by

$$G=(\mathcal{V},\mathcal{E}),$$

where $\mathcal{V}=\Sigma^{*}$ and the set $\mathcal{E}$ of labeled edges is partitioned into three subsets:

\begin{itemize}
 \item $\mathcal{E}_{1}=\{(U,1,U1),\mbox{ solid edges}\}$
 \item $\mathcal{E}_{2}=\{(U,2,U2),\mbox{ solid edges}\}$
\item $\mathcal{E}_{0}=\{(U,0,\Theta(U)2),\mbox{ dashed edges}\}$
\end{itemize}

Hence, the graph $G$ is obtained by adding to an infinite complete binary tree (with edge labels in $\Sigma$ and node labels in $\Sigma^{*}$) one additional edge outgoing from each non-empty node $U$, labeled by $0$, and ingoing to the node $\Theta(U)2$.

\begin{figure}
\begin{center}
\includegraphics[height=180mm]{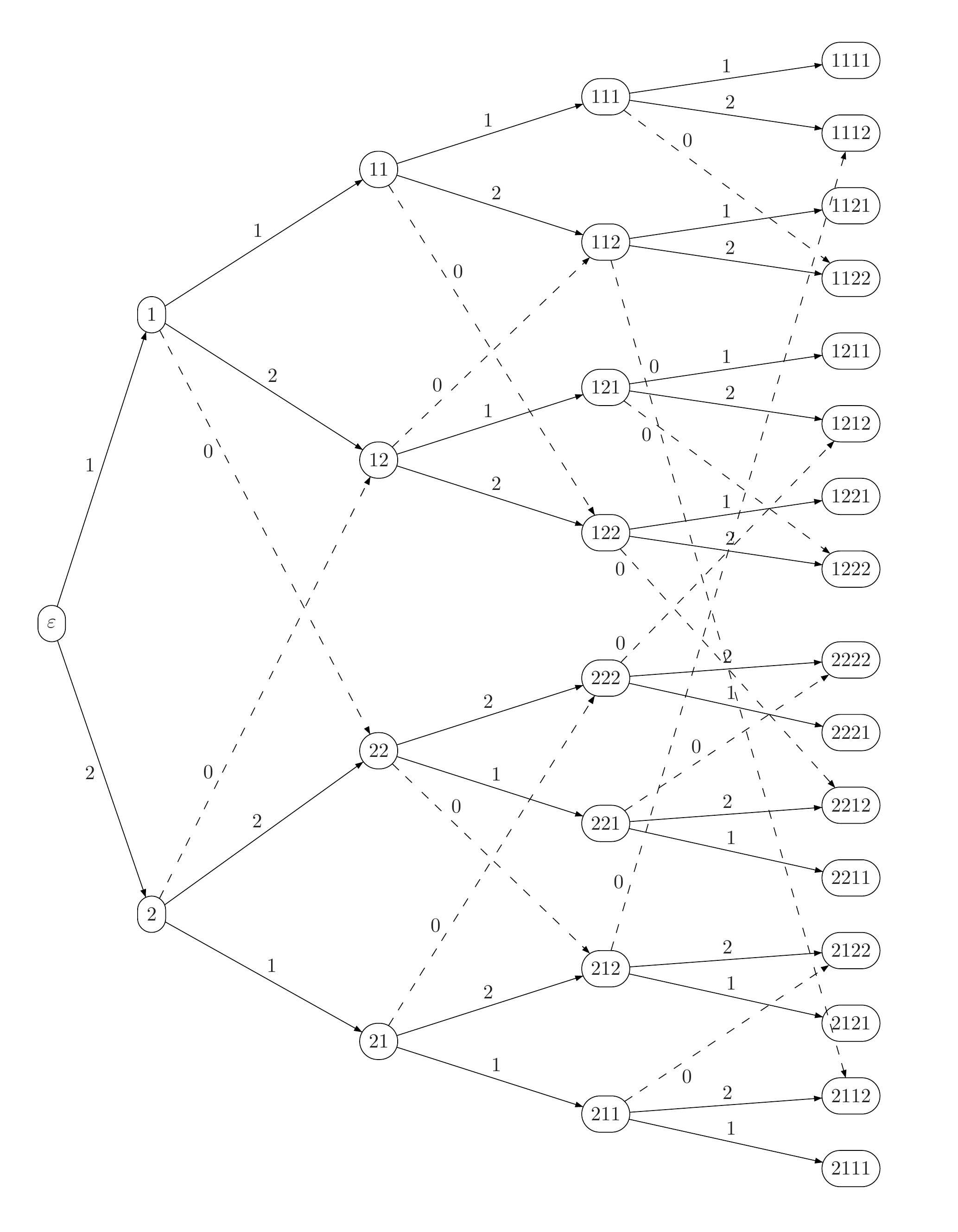}
\caption{The graph $G$ cut at height $4$.\label{fig:UVCA}}
\end{center}
\end{figure}

A partial diagram of the graph $G$ is depicted in \figref{fig:UVCA}. The edges in $\mathcal{E}_{0}$ are dashed.

\begin{figure}
\begin{center}
\includegraphics[height=70mm]{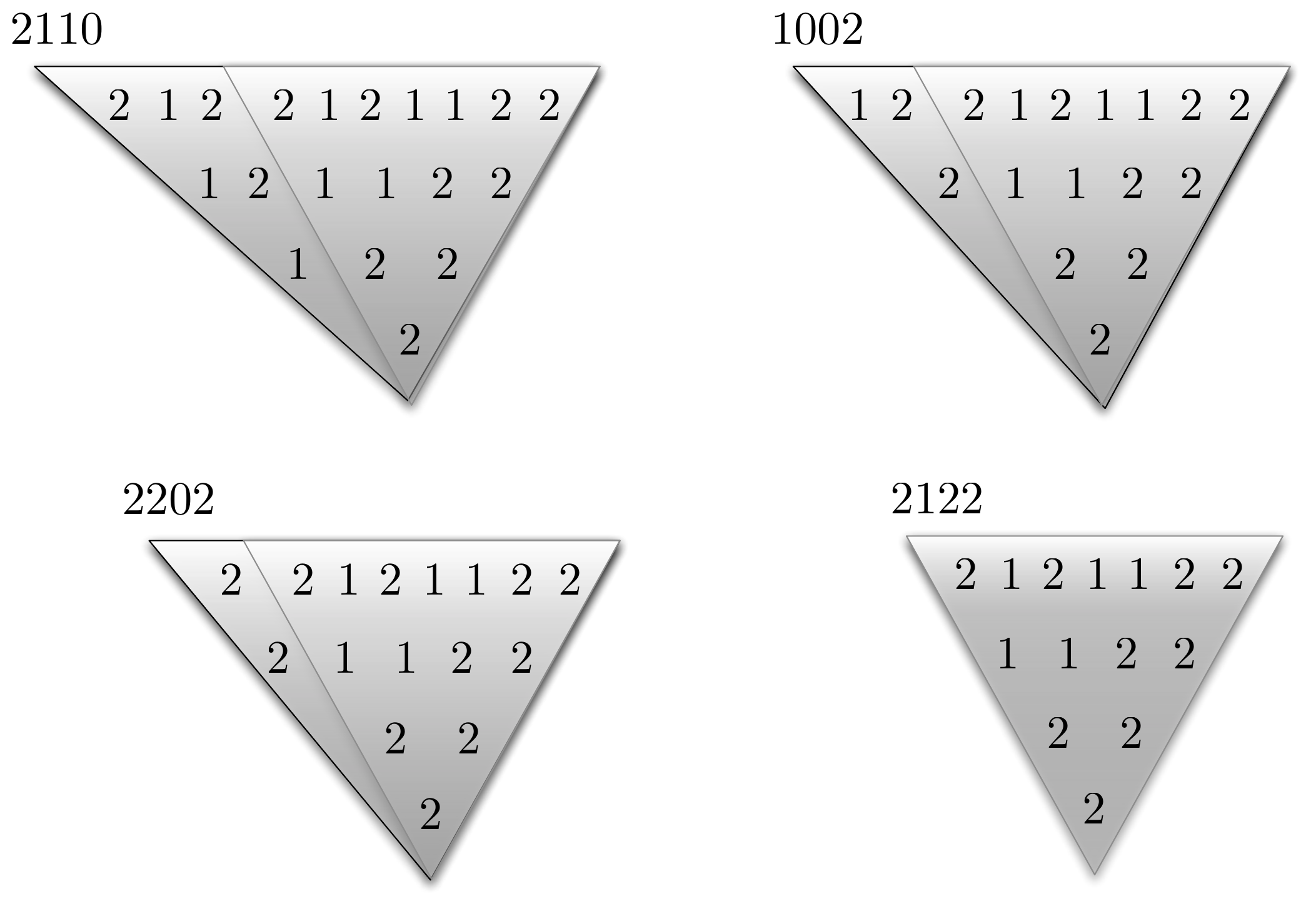}
\caption{The paths in $G$ starting at the origin and ending in the node $2122$ (left frontier of the single-rooted minimal word $u=2121122$) are: $2110$, $1002$, $2202$ and $2122$ (to see this, follow the paths labeled by these words in \figref{fig:UVCA}). These are precisely the $\Psi$-classes forming the minimal class $2122$, and are the left frontiers of the \C-words $212\cdot u$, $12 \cdot u$, $2 \cdot u$ and $u$, respectively.\label{fig:Class}}
\end{center}
\end{figure}

So, the  nodes of $G$ represent the minimal classes, and the $\Psi$-classes forming the minimal class $U\in \Sigma^{*}$ are the labels of the paths starting at $\epsilon$ and ending in $U$. An example is given in  \figref{fig:Class}.

In \cite{FeFi12}, we used the structure of the graph $G$ to prove the following result:

\begin{theorem}
  Let $u\in \C$. Then, there exists $z$ such that $uzu\in \C$ and $|uzu| \le C|u|^{2.72}$, for a suitable constant $C$.
\end{theorem}

The theorem above is a weak version of Conjecture \ref{conj:univ}. The validity of the following conjecture on the graph $G$ would imply the validity of Conjecture \ref{conj:univ}.

\begin{conjecture}\label{conj:univG}
 Let $U,V\in \Sigma^{*}$. Then, there exists $Z\in \Sigma^{*}$ and paths in $G$ from $U$ to $Z$ and from $V$ to $Z$.
\end{conjecture}

Indeed, it is easy to see that Conjecture \ref{conj:univ} is true if the following holds: For any \C-words $u$ and $v$, there exist $u'$ and $v'$ such that $uu'$ and $vv'$ are \C-words belonging to the same minimal class. And the latter condition holds if Conjecture \ref{conj:univG} is true.

We also state the following.

\begin{conjecture}\label{conj:linearG}
There exists a linear integer function $f$ such that for every $k\ge 0$, given two words $U\in \Sigma^{k}$ and $Z\in \Sigma^{f(k)}$, there exists in $G$ a path from $U$ to $Z$. 
\end{conjecture}

The validity of Conjecture \ref{conj:linearG} would imply that any (double-rooted maximal) \C-word of height $f(k)$ would contain all the \C-words of height $k$ as factors. In particular, this would imply that any right-infinite \C-word (and so, in particular, the Kolakoski word) is uniformly recurrent. 

\section{Recursivity}\label{sec:recursive}

In this section we prove that the maps $\GammaS$ and $\GammaD$, and therefore the map $\Theta$, can be defined recursively and with no explicit reference to \C-words. This also leads to a recursive definition of the graph $G$, since the graph $G$ is completely defined by the action of $\Theta$. The base case can be set for the three maps as follows: $\GammaS(\epsilon)=\GammaD(\epsilon)=\Theta(\epsilon)=\epsilon$.

For any vertical word $U\in \Sigma^{*}$, we will denote by $\kappa(U)$ the complement of $U$. As a consequence of Proposition \ref{prop:front}, we have:

\begin{lemma}
 For any $U\in \Sigma^{*}$, $\kappa \GammaD (U)= \GammaD \kappa (U)$.
\end{lemma}

In the following theorem we give recursive formulae for $\GammaS$ and $\GammaD$. 

\begin{theorem}\label{theor:recursive}
For any $U\in \Sigma^{*}$, the following recursive formulae hold:
\begin{enumerate}
 \item $\GammaS (U1)=\GammaD  \GammaS  \GammaD(U)1$;
 \item $\GammaS (U2)=\kappa \GammaD(U)2$;
 
 \item $\GammaD (U1)=\kappa \GammaD  \GammaS  \GammaD  \GammaS  \GammaD (U)2$;
 \item $\GammaD (U2)= \GammaD  \GammaS  \GammaD  \GammaS  \GammaD \kappa (U)1$.
\end{enumerate}
\end{theorem}

\begin{figure}[ht]
\begin{center}
\includegraphics[height=47mm]{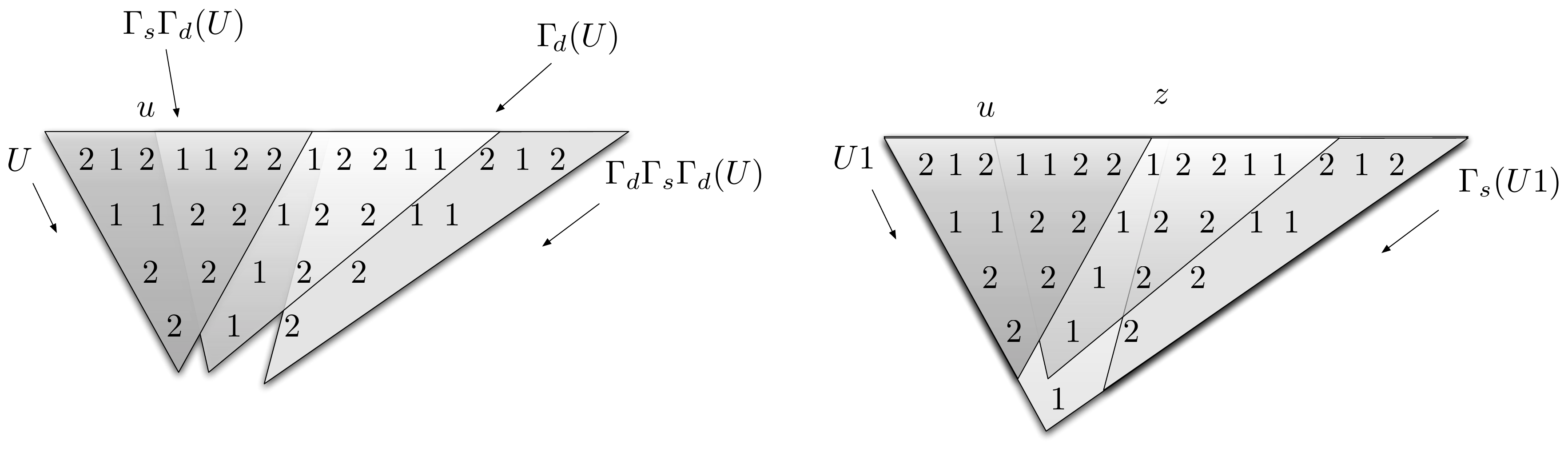}
\caption{The proof of the first recursion of Theorem \ref{theor:recursive}. The single-rooted minimal word $z=212112212211212$ (in lighter gray, on the right) has left frontier $U1=21221$, where $U$ is the left frontier of the single-rooted minimal word $u=2121122$ (in darker gray, on the left). The right frontier of $z$ is $\GammaS(U1)=\GammaD\GammaS\GammaD(U)1=21221$. \label{fig:Proof}}
\end{center}
\end{figure}

\begin{proof}
1. Let $U\in \Sigma^{*}$. Consider the single-rooted minimal word $z$ whose left frontier is $U1$. Generalizing the example illustrated in \figref{fig:Proof}, it is easy to see that the right frontier of $z$ is $\GammaD\Theta(U)1$, i.e., $\GammaD\GammaS\GammaD(U)1$.

2. Follows from the first part of Proposition \ref{prop:front}.

3. Let $U\in \Sigma^{*}$. Consider the double-rooted minimal word $z'$ whose left frontier is $U1$. From 1. and 2., it follows that the right frontier of $z'$  is the complement of the right frontier of the word having left frontier $\GammaD  \GammaS  \GammaD  \GammaS  \GammaD(U)$ with $2$ appended.

4. Symmetric to 3.
\end{proof}

\begin{remark}
The statement of Theorem \ref{theor:Theta} can also be proved using the recursive formulae of Theorem \ref{theor:recursive}. Indeed, in order to prove the equality $\GammaS(U0)=\GammaS(\Theta(U)2)$, it is sufficient to prove it on the complements. By Proposition \ref{prop:front}, we have $\kappa\GammaS(U0)=\GammaS(U1)$. Always by Proposition \ref{prop:front}, we have $\kappa\GammaS(\Theta(U)2)=\GammaD\Theta(U)1=\GammaD\GammaS\GammaD(U)1$, and the statement then follows from the first recursion of Theorem \ref{theor:recursive}.
\end{remark}

From the recursive formulae for $\GammaS$ and $\GammaD$ we can derive recursive formulae for $\Theta$, using standard algebraic manipulations.

\begin{corollary}\label{cor:piteta}
The following recursive formulae hold:
  \begin{enumerate}
  \item $\Theta (U1) = \Theta^{2}(U)2$;\label{theta1}
 \item $\Theta (U2) = \Theta\kappa(U)1$.\label{theta2}
\end{enumerate}
\end{corollary}

In conclusion, we have proved the following result.

\begin{theorem}\label{theor:mainrecursion}
The graph $G$ is obtained by adding to the infinite complete binary tree over $\{1,2\}$ the additional  edges $(U,0,\Theta(U)2)$, for each $U\in \Sigma^{*}\setminus\{\epsilon\}$, defined recursively by the formulae in Corollary \ref{cor:piteta}.
\end{theorem}

That is, the graph $G$ can be defined recursively and with no explicit reference to $\C$-words.

\section{Conclusions and open problems}\label{sec:final}

The vertical representation is a compact representation of $\C$-words that allows one to represent any $\C$-word of length $n$ by means of two words whose length is logarithmic in $n$ (the frontiers). We defined two maps, $\GammaS$ and $\GammaD$,  acting on the frontiers, that allowed us to introduce an equivalence relation (the minimal equivalence) and  reduce the study of \C-words to the equivalence classes of this relation (the minimal classes). These classes can be represented over an infinite directed acyclic graph $G$, that is completely defined by the action of the map $\Theta$, which is the composition of $\GammaS$ and $\GammaD$. We proved that $\Theta$, and therefore $G$, can be defined recursively and independently from the context of $\C$-words.

Besides being more compact, we believe that the new representation presented here will allow the use of results from graph theory or poset theory in the study of \C-words. 
As an illustration, we formulated two new conjectures (Conjecture \ref{conj:univG} and Conjecture \ref{conj:linearG}) on the graph $G$ that, if proven, would imply the validity of important conjectures on $\C$-words. In particular, Conjecture \ref{conj:univG} states that the partial order on the set $\Sigma^{*}$ of minimal classes defined by: 
``$U \le V$ if and only if there exists in $G$ a path from $U$ to $V$'',
makes $\Sigma^{*}$ a \emph{directed set}, i.e., a set in which every pair of elements has an upper bound. 
Notice that this relation does not make $\Sigma^{*}$ a \emph{lattice}, since the least upper bound between two elements is not always defined (e.g., the minimal classes $12$ and $11$ have the two upper bounds $112$ and $122$, see \figref{fig:UVCA}).

\section{Acknowledgements}\label{sec:ack}

We warmly thank an anonymous referee whose comments helped us to highlight  the attribution of some definitions and conjectures to R.~Oldenburger.


\end{document}